\documentclass[runningheads]{llncs}
\usepackage{graphicx}
\usepackage[colorlinks=true,linkcolor=blue]{hyperref}

\usepackage{amsfonts}
\usepackage{amsmath}
\DeclareMathOperator*{\argmin}{arg\,min}
\usepackage{array}
\usepackage{multirow}

\begin{document}
\title{Implicit field learning for unsupervised anomaly detection in medical images}
%
%
\author{Sergio {Naval Marimont}\inst{1} \and
Giacomo Tarroni\inst{1, 2}}
%


\authorrunning{S. Naval Marimont and G. Tarroni}
%
\institute{CitAI Research Centre, City, University of London, London, UK \and
BioMedIA, Imperial College, London, UK
\email{\{sergio.naval-marimont,giacomo.tarroni\}@city.ac.uk}}
\maketitle              
\begin{abstract}
We propose a novel unsupervised out-of-distribution detection method for medical images based on implicit fields image representations. In our approach, an auto-decoder feed-forward neural network learns the distribution of healthy images in the form of a mapping between spatial coordinates and probabilities over a proxy for tissue types. At inference time, the learnt distribution is used to retrieve, from a given test image, a restoration, i.e. an image maximally consistent with the input one but belonging to the healthy distribution. Anomalies are localized using the voxel-wise probability predicted by our model for the restored image. We tested our approach in the task of unsupervised localization of gliomas on brain MR images and compared it to several other VAE-based anomaly detection methods. Results show that the proposed technique substantially outperforms them (average DICE 0.640 vs 0.518 for the best performing VAE-based alternative) while also requiring considerably less computing time.

\end{abstract}
\section{Introduction}
Multiple deep learning methods have been proposed to automatically localize anomalies in medical images, with fully-supervised approaches being able to achieve high segmentation accuracies \cite{Litjens2017}. However, these methods 1) rely on the availability of large and diverse annotated datasets for training, and 2) they are specific to the anomalies annotated in the dataset and are therefore unable to generalize to previously unseen pathologies. On the other hand, the unsupervised learning paradigm is not affected by these limitations. Unsupervised approaches usually aim at learning the distribution of healthy/normal unannotated images and at classifying as anomalies the images that differ from the learnt distribution. Two categories of generative models, namely Variational Auto-Encoders (VAEs) and Generative Adversarial Networks (GANs), have been implemented in many techniques for unsupervised anomaly detection. However, comparative studies \cite{baur2021autoencoders} show that their performance is still far from that of equivalent supervised methods. 

\subsubsection{Related Works:} Generally, anomaly detection techniques make use of generative models to learn the distribution of healthy/normal images and leverage the learnt distribution to compute voxel-by-voxel anomaly scores (AS), which identify image areas that differ from the normal anatomy. The vanilla VAE-based approach assumes that a model trained on normal data will not be able to reconstruct anomalies, and consequently that the voxel-wise reconstruction loss can be used as AS. However, auto-encoders with enough capacity can also reconstruct abnormal samples, making VAE reconstruction loss a poor AS \cite{baur2021autoencoders,Zimmerer2019}. Several methods have tried to over-come this limitation. For instance, \cite{Zimmerer2019} proposed to leverage the KL divergence gradient w.r.t voxels as AS. \cite{zimmerer2018context} added context-encoding tasks to incentivise the VAE to properly generate restored (i.e. anomaly-free) images. In \cite{Chen2020}, the authors proposed to restore images by minimising a loss function composed by the VAE ELBO and a data consistency term. \cite{marimont2021anomaly} proposed to generate restorations with a vector-quantized VAE by resampling low-probability latent variables. GAN-based approaches \cite{schlegl2017unsupervised,Schlegl2019}, rely on a similar ideas for restoration. However, GANs notably suffer from mode collapse, i.e. the tendency to learn to generate samples only from a subset of the normal image distribution. In addition, also GANs can generate anomalous samples \cite{baur2021autoencoders}. Due to these issues, most of the approaches based on generative models have yielded limited anomaly detection performance, struggling to reach DICE scores of 0.5 in brain MRI datasets \cite{baur2021autoencoders}.

Techniques based on supervised learning using datasets with synthetically-generated anomalies \cite{Tan2020} have been very recently proposed for anomaly detection. They have achieved high accuracy in the MICCAI 2020 Medical Out-of-Distribution challenge (MOOD) \cite{moodchallenge2020}, which partially included synthetic anomalies in its test set. While promising, these approaches move the focus to the task of generating realistic anomalies, and their performance on datasets with real abnormalities remains largely unexplored.

\subsubsection{Contributions:} Recently, an approach referred to as implicit field learning (or occupancy net-works) has been introduced to reconstruct 3D shapes through learning their implicit surface representation \cite{mescheder2019occupancy,chen2019learning}. Instead of using convolutional networks to learn a distribution over a dense set of voxels, in this approach a linear neural net-work learns to map continuous spatial coordinates to either object/background labels (binary classification) \cite{mescheder2019occupancy,chen2019learning} or to the signed distance function with respect to the object surface \cite{park2019deepsdf}. Importantly, the authors of \cite{park2019deepsdf} also proposed to substitute the auto-encoder architecture with an auto-decoder architecture, removing the need for an encoder network to obtain the latent representation.

In this paper, we propose a novel approach to unsupervised anomaly detection that leverages the implicit field learning paradigm. Our main contributions are the following:
\begin{itemize}
    \item We propose a modification of the implicit field learning technique that enables learning relevant anatomical features for unsupervised anomaly detection;
    \item We propose an anomaly detection neural network pipeline which overcomes the limitations of VAE models: by relying on an auto-decoder architecture, our network generates anomaly-free reconstructions. Additionally, the implicit field representation is detached from a specific input resolution and can be scaled seamlessly to deal with high resolution 3D medical images;
    \item We tested this approach in the task of unsupervised localization of gliomas on brain MR images and compared it to several other VAE-based approaches. The proposed technique substantially outperforms the competition both in terms of accuracy (average DICE 0.640 vs 0.518 for the best performing competitor) and computing speed (55.4 s vs 293.1 s, respectively).
\end{itemize}

\section{Methods}
\subsubsection{Implicit field representation:} While a 3D image is typically represented by the intensities of a set of discrete voxels, implicit field networks learn a continuous function $f$ with spatial coordinates $\mathbf{p}=(x,y,z) \in \mathbb{R}^3$ as input. Instead of a binary label for object/background classification, we propose that this function maps to the probability distribution over $C$ classes, each representing a range of voxel intensities (see next section). In addition to the spatial coordinates, the network receives as input a latent variable $\mathbf{z} \in \mathbb{R}^D$  which describes a specific 3D image:

\begin{equation} \label{eq:1}
    f: \mathbb{R}^3 \times \mathbb{R}^D \to \{1,2,...C\}
\end{equation} 

The network $f$ learns the posterior probability over intensity ranges for continuous spatial coordinates $\mathbf{p}$ and latent variables $\mathbf{z}$. We model the posterior probability using a \textit{softmax} activation:

\begin{equation} \label{eq:2}
    P(t=j \mid \mathbf{z,p}) = \frac{\exp f^j(\mathbf{z,p})}{\sum^C_i{\exp f^i (\mathbf{z,p})}} 
\end{equation} 

\begin{figure}
\includegraphics[width=\textwidth]{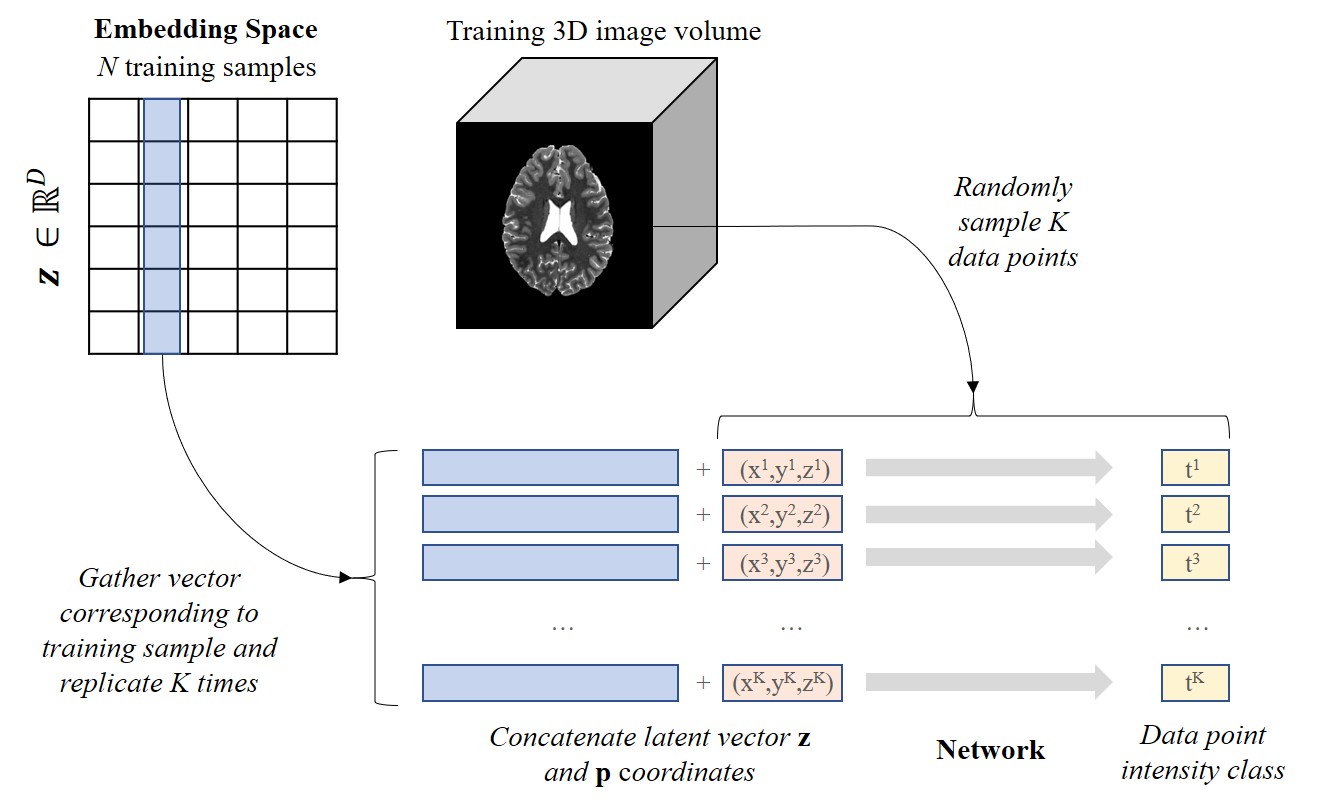}
\caption{Schema for auto-decoder training. A latent vector is obtained for a training sample and concatenated with coordinates sampled from the 3D image volume. The network learns the mapping (latent features, coordinates) $\to$ multiclass encoding of voxels intensities.} \label{fig1}
\end{figure}

The latent variables $\mathbf{z}$ are obtained using the auto-decoder architecture proposed in \cite{park2019deepsdf}: as opposed to training an encoder network to produce the latent representation, in the auto-decoder approach each training 3D image is paired with a D-dimensional vector in an embedding space. During training, backpropagation optimizes not only the network parameters but also the latent vector representation of each 3D image. In the auto-decoder architecture (see Fig. \ref{fig1}), at inference time, the latent vector is initialized randomly and optimization is used to find the latent representation that better represents the test 3D image.

Specifically, during training the expression \ref{eq:3} is used to optimize latent codes and parameters $\mathbf{\theta}$ of network $f_{\theta}$ by sampling $K$ data points from $N$ training 3D images:

\begin{equation} \label{eq:3}
    \argmin_{\theta,\{\mathbf{z}_i\}^N_{i=1}}
    \sum^N_{i=1}{(\sum^K_{j=1}{\mathcal{L}(f_\theta(\mathbf{z}_i,\mathbf{p}_{i,j}),t_{i,j})+\frac{1}{\sigma^2}\lVert\mathbf{z}_i\rVert^2_2})}
\end{equation}

Note that $\mathcal{L}$ is the cross-entropy loss between network output and the true voxel class $t \in \{1,2,…C\}$. Similarly to \cite{park2019deepsdf}, we assumed the prior for the latent codes distribution to be a spherical multivariate-Gaussian with covariance $\sigma^2 I$. The $\sigma$ hyperparameter allows to modulate the amount of regularization in the latent distribution. During inference, the expression \ref{eq:3} is optimized only for $\mathbf{z}$ (fixing network parameters $\theta$), obtaining the latent code that best describes a given test 3D image.

\begin{figure}[b!]
\includegraphics[width=\textwidth]{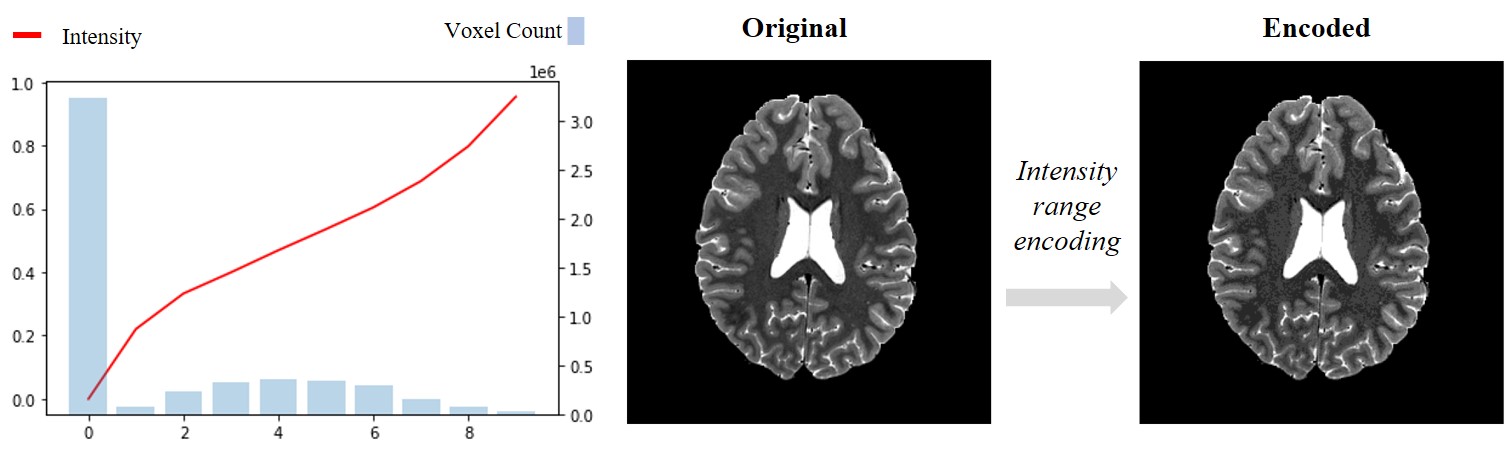}
\caption{Left: intensity and count of voxels per cluster (using kMeans, with k = 10, on 2 million voxels randomly sampled across multiple subjects). Right: effects of intensity range encoding on a sample image.} \label{fig2}
\end{figure}

We also utilized the coordinate encoding function described in \cite{mildenhall2020nerf}, by which each of the coordinates in $\mathbf{p}$ is first normalized to the range [-1,1] and later encoded using the expression \ref{eq:4} (shown for x. We used L=10 in our experiments):
\begin{equation} \label{eq:4}
    \gamma(x)=(\sin(2^0 \pi x),\cos(2^0 \pi x),...,\sin(2^{L-1} \pi x),\cos(2^{L-1} \pi x))
\end{equation}	

\subsubsection{Intensity range encoding:} Instead of modelling 3D image intensities as a continuous variable, we discretize the intensity values in $C$ clusters, allowing the neural network to learn the probability distribution over $C$ classes of intensities. We wish to define the intensity clusters so that the encoding preserves as much information as possible in the original volume. With this objective, we use k-Means, treating the number of clusters as an hyper-parameter (see Fig. \ref{fig2}).

The rationale behind intensity range encoding is two fold. First, it enables a rather straightforward extension of implicit field learning approaches to the task of image reconstruction. Second, it allows to untie the computation of the AS from distances in image intensities between original image and reconstructed version (which is the most common strategy for approaches based on generative models). Instead, in the proposed technique the AS is derived from the predicted probability over intensity clusters, which we assume to better represent different tissue types.

\subsubsection{Mode-pooling smoothing:} We found that smoothing and denoising the 3D images volumes slightly improved anomaly detection accuracy for our approach. In order to preserve the overall structures and only remove spurious intensity values, we propose a 3D \textit{mode-pooling} layer which, for a 3-dimensional sliding window, returns the most common intensity cluster. We used a 2x2x2 \textit{mode-pooling} filter in our training set and 3x3x3 in validation and test sets.

\subsubsection{Voxel-wise Anomaly Score (AS):} At inference time, we aim at retrieving a healthy image from the model consistent with a test image. The retrieved image is called a \textit{restoration}, as it preserves consistency with the test image but it belongs to the learnt distribution of healthy images. Anomalies are finally located by comparing the restoration with the test image. In order to generate a restoration, we move along latent space searching for the latent vector $\mathbf{z}$ that minimizes the following expression over $K$ randomly sampled data points:
\begin{equation} \label{eq:5}
    \argmin_{\mathbf{z}}
    (\sum^K_{j=1}{\mathcal{L}(f_\theta(\mathbf{z},\mathbf{p}_j),t_j)+\frac{1}{\sigma^2}\lVert\mathbf{z}\rVert^2_2})
\end{equation}

Minimization is performed with Adam optimizer for 700 steps with $K$ = 16,200. Once a restoration is generated with the retrieved $\mathbf{z}$, we can compute a voxel-based anomaly score (AS). Specifically, we estimate the probability over intensity clusters for each voxel using the network and compute the voxel-wise cross-entropy loss between the test image and the restoration as anomaly score:
\begin{equation} \label{eq:6}
    AS = -\log P(t=t_{GT} \mid \mathbf{z,p})
\end{equation}
	
where $t_{GT}$ is the true voxel intensity after intensity range encoding. The proposed voxel-wise AS is similar to the one in \cite{schlegl2017unsupervised}, replacing absolute residuals with cross-entropy to account for the intensity range encoding. We also perform post-processing to denoise the obtained AS obtained using a min-pooling layer (with filter size = 3) and average-pooling layer (with filter size = 15), both 3-dimensional.

\section{Experiments and Results}
\subsubsection{Experimental set-up:} We tested our approach by training the proposed technique on a dataset of brain MR images from healthy subjects and testing it on images with gliomas. As benchmarks, we trained and tested 3 VAE-based anomaly detection techniques. Since these methods have been originally presented with 2D architectures, we created a 2D version of our approach (which used MR slices as inputs) to enable a fairer comparison. In the 2D experiments we processed 1 every 4 axial slices, (i.e., 40 slices per volume). We then evaluated our approach in its native 3D implementation using 3D MR image volumes for training and testing. 

\subsubsection{Datasets and data pre-processing:} We use two publicly available brain MRI datasets:
\begin{itemize}
    \item The Human Connectome Project Young Adult (HCP) dataset \cite{van2012human} with images of 1,112 young and healthy subjects.
    \item The Multimodal Brain Tumor Image Segmentation Benchmark (BRATS) \cite{menze2014multimodal}, 2018 edition dataset, consisting of images with annotated gliomas.
\end{itemize}
The training set consists of 1,055 images from HCP, the test set of 50 images randomly sampled from BRATS  and the validation set of 11 (6 from BRATS and 5 from HCP), used for hyper-parameter tuning. In both HCP and BRATS we use the pre-processed, skull-stripped T2-weighted structural images. Additionally, in all experiments but one, both datasets were downsampled to 160x160x160 resolution, intensities were clipped to the percentile 98 and later normalized to the range [0,1]. In one experiment, we tested our approach using instead the original, high-resolution images, training at 300 voxel resolution in HCP and testing at 240 voxel resolution in BRATS. No augmentations were performed in training the proposed technique. Elastic transforms, scaling, rotations and random brightness and contrast were instead applied to all VAEs benchmark experiments. In VAE experiments, images are also normalized to have zero mean and unit standard deviation.

For our approach, k=10 was chosen for intensity clustering after tuning.

\subsubsection{Network architecture and implementation details:} We used the same network architecture and training details from \cite{park2019deepsdf} for all our experiments. The decoder is a feed-forward network composed of 8 fully-connected layers. Latent dimensionality is 256, all hidden layers have 512 units and use ReLU as activation and weight normalization. We apply dropout in all layers with probability 0.2. The embedding space is initialized with $N(0, 0.01^2)$ and the prior covariance hyper-parameter is set to $\sigma=0.01$. Training lasted 2,000 epochs with Adam optimizer and we applied a learning rate decay. Training batches are composed of 97,200 randomly sampled points (16,200 points from 6 different volumes). Implementation, trained models and test sets identifiers are made publicly available in \footnote{\href{https://github.com/snavalm/ifl_unsup_anom_det}{\texttt{https://github.com/snavalm/ifl\_unsup\_anom\_det}}}. All experiments were run using a Nvidia GTX 1070 GPU.

In 2D experiments, we assign a latent code with 256 dimensions to each axial slice instead of the whole volume and prediction is also conditioned on the axial coordinate. At inference time, we obtain an AS for one axial slice every four. Each axial AS is replicated 4 times to return to the original axial resolution and the AS post-processing (min-pooling and average-pooling) is performed with 3D filters on the resulting volume. We also followed this methodology in all 2D VAE benchmarks.

\begin{table}[b!]
\caption{Experimental results on BRATS 2018 dataset.}\label{Table:tab1}
\begin{tabular}{p{3cm} >{\centering\arraybackslash}p{1.2cm} >{\centering\arraybackslash}p{2.2cm} >{\centering\arraybackslash}p{1cm} >{\centering\arraybackslash}p{1.3cm} >{\centering\arraybackslash}p{1.5cm} >{\centering\arraybackslash}p{1.3cm}}
\hline
Method &  [DICE] & DICE ($\mu \pm \sigma$) & AP & AUROC & FPR@95R & IT (s) \\
\hline
\textbf{2 dimensional} \\
\hline
VAE $^{(1)}$ & 0.472 & 0.447 $\pm$ 0.161 & 0.477 & 0.949 & 0.2229 & 0.1 \\
VAE restoration $^{(2)}$ & 0.417 & 0.390 $\pm$ 0.146 & 0.413 & 0.936 & 0.2448 & 79.1 \\
VQ-VAE $^{(3)}$ & 0.568 & 0.518 $\pm$ 0.188 & 0.593 & 0.972 & 0.1366 & 293.1 \\
IF 2D (ours) & 0.612 & 0.555 $\pm$ 0.178 & 0.665 & 0.991 & 0.0456 & 55.4 \\
\hline
\textbf{3 dimensional} \\
\hline
IF 3D (ours) & 0.681 & 0.640 $\pm$ 0.177 & 0.733 & 0.992 & 0.0462 & 51.1 \\
IF 3D* (ours) & \textbf{0.716} & \textbf{0.672 $\pm$ 0.155} & \textbf{0.771} & \textbf{0.994} & \textbf{0.0386} & 64.1 \\
\hline
\multicolumn{7}{l}{1 - VAE with 10 latent dimensions, L1 reconstruction loss.} \\
\multicolumn{7}{l}{2 - VAE with 128 latent dimensions, 500 restoration steps as described in \cite{Chen2020}.} \\
\multicolumn{7}{l}{3 - VQ-VAE 20x20 latent, 8 restorations. Implementation and processing from \cite{marimont2021anomaly}.} \\
\multicolumn{7}{l}{* - Train and test in original high-resolution (300 voxel HCP and 240 BRATS)} \\
\end{tabular}
\end{table}

\subsubsection{Performance evaluation:} In order to assess the voxel-wise anomaly detection performance we followed the conventions used in \cite{baur2021autoencoders}. We report the best possible DICE score [DICE] in our test set which is calculated as the maximum DICE taking into consideration all individual voxels in the test set. Additionally, we determine the optimal threshold for AS using the validation set and calculate the DICE score using this threshold for each subject in the test set. Mean and standard deviations of subject specific DICE scores are reported in \textit{\textbf{Table \ref{Table:tab1}}}. We also report Average Precision (AP), area under Receiver Operating Characteristics (AUROC), the False Positive Rate at 95\% recall (FPR@95R) and inference time per image volume in seconds (IT (s)).

\begin{figure}[b!]
\includegraphics[width=\textwidth]{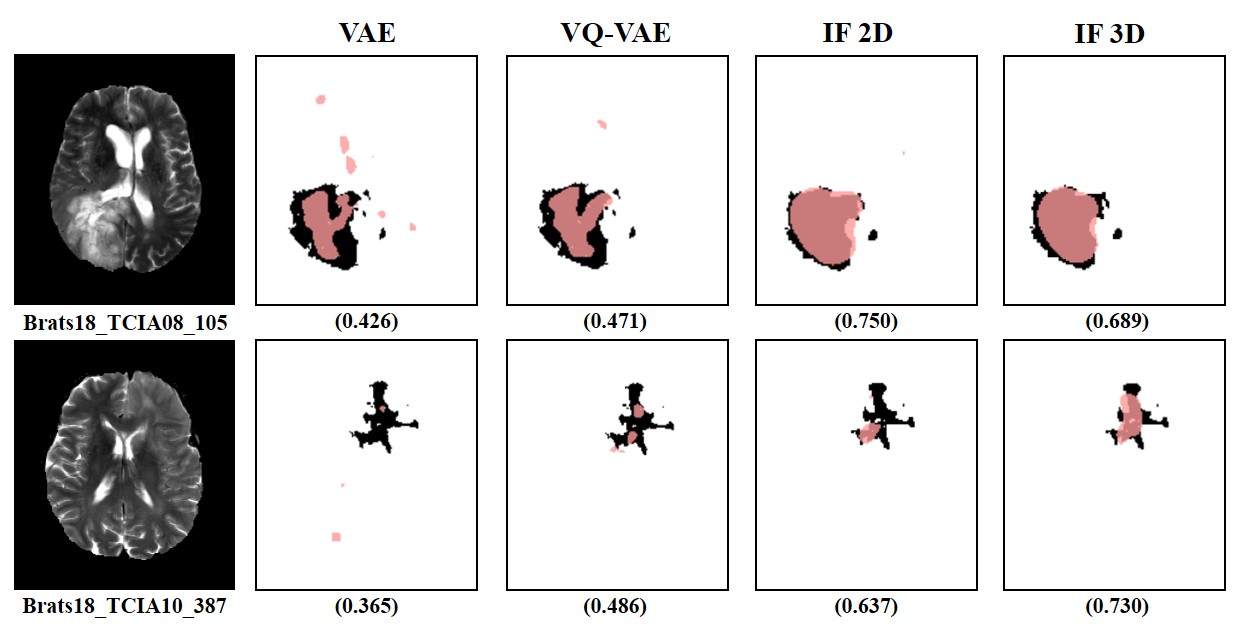}
\caption{Visual comparison of anomaly scores (pink) versus ground truth (black) for two different subjects (central axial slice). In brackets, DICE score for the whole subject.} \label{fig3}
\end{figure}

At 160 voxel resolution, the proposed 3D implicit fields (IF) method improved mean DICE score by 12 points (0.640 vs 0.518 of VQ-VAE). A moderate increase of 4 points was also observed when comparing to the 2D IF implementation. Importantly, our method can also seamlessly capitalize, without architectural modifications, on the 3D high resolution images, with the mean DICE score increasing to 0.672. While convolutional architectures require pre-specified resolutions, the implicit fields auto-decoder approach allowed us to train and test in different resolutions. Qualitative results are shown in Fig. \ref{fig3}. 

These results show that our method improves on the state of the art in the task of localizing anomalies when training and testing using different datasets with different acquisition protocols. It is expected that image augmentation would help alleviate the differences produced by diverse acquisition pipelines, however augmentations were not applied in our method implementation. The auto-decoder architecture forces a latent code to be pre-assigned to each training sample, consequently the augmented images need to be consistent across training epochs. 

Inference computing time is also moderate, with 51 seconds per volume for the 3D experiments. Note that for 2D methods, the time reported correspond to only one every four axial slices and consequently it is not directly comparable with IF 3D methods. The inference time is mostly associated with the optimization required to retrieve the latent test image representation. Adding an encoder network to the architecture would further reduce the inference time (either switching to an auto-encoder architecture or training an encoder after the auto-decoder, similarly to \cite{Schlegl2019}). 

The optimization used to retrieve restorations translates in stochasticity at inference time because the minimization can converge to slightly different restorations. This could become a challenge if the algorithm converged to bad local minima, however it may present as well an opportunity to improve results by calculating the AS taking into consideration multiple restorations near the optimal $\mathbf{z}$.

Finally, in our experiment VAE restoration underperformed VAE reconstruction loss, which is unexpected given the previous studies \cite{baur2021autoencoders,Chen2020}. Differences in our implementation, namely deeper architectures with residual blocks, a smaller latent space, batch normalization and differences in image normalization, could have improved our VAE reconstruction or limited the effectivity of the restoration method.

\section{Conclusion}

We presented a novel unsupervised anomaly segmentation method based on implicit field learning that outperforms previous VAE-based approaches in glioma segmentation in brain MR images. In the future, we intend to perform further evaluations relative to other brain pathologies and medical image modalities. 

%
%

%
%
%
\bibliographystyle{splncs04}
\bibliography{ref.bib}

\begin{thebibliography}{10}
\providecommand{\url}[1]{\texttt{#1}}
\providecommand{\urlprefix}{URL }
\providecommand{\doi}[1]{https://doi.org/#1}

\bibitem{baur2021autoencoders}
Baur, C., Denner, S., Wiestler, B., Navab, N., Albarqouni, S.: Autoencoders for
  unsupervised anomaly segmentation in brain mr images: A comparative study.
  Medical Image Analysis p. 101952 (2021)

\bibitem{Chen2020}
Chen, X., You, S., Tezcan, K.C., Konukoglu, E.: Unsupervised lesion detection
  via image restoration with a normative prior. Proceedings of The 2nd
  International Conference on Medical Imaging with Deep Learning  \textbf{PMLR
  102},  540--556 (2020)

\bibitem{chen2019learning}
Chen, Z., Zhang, H.: Learning implicit fields for generative shape modeling.
  In: Proceedings of the IEEE/CVF Conference on Computer Vision and Pattern
  Recognition. pp. 5939--5948 (2019)

\bibitem{Litjens2017}
Litjens, G., Kooi, T., Bejnordi, B.E., Setio, A.A.A., Ciompi, F., Ghafoorian,
  M., Van Der~Laak, J.A., Van~Ginneken, B., S{\'a}nchez, C.I.: A survey on deep
  learning in medical image analysis. Medical image analysis  \textbf{42},
  60--88 (2017)

\bibitem{menze2014multimodal}
Menze, B.H., Jakab, A., Bauer, S., Kalpathy-Cramer, J., Farahani, K., Kirby,
  J., Burren, Y., Porz, N., Slotboom, J., Wiest, R., et~al.: The multimodal
  brain tumor image segmentation benchmark (brats). IEEE transactions on
  medical imaging  \textbf{34}(10),  1993--2024 (2014)

\bibitem{mescheder2019occupancy}
Mescheder, L., Oechsle, M., Niemeyer, M., Nowozin, S., Geiger, A.: Occupancy
  networks: Learning 3d reconstruction in function space. In: Proceedings of
  the IEEE/CVF Conference on Computer Vision and Pattern Recognition. pp.
  4460--4470 (2019)

\bibitem{mildenhall2020nerf}
Mildenhall, B., Srinivasan, P.P., Tancik, M., Barron, J.T., Ramamoorthi, R.,
  Ng, R.: Nerf: Representing scenes as neural radiance fields for view
  synthesis. In: European Conference on Computer Vision. pp. 405--421. Springer
  (2020)

\bibitem{marimont2021anomaly}
Naval~Marimont, S., Tarroni, G.: Anomaly detection through latent space
  restoration using vector quantized variational autoencoders. In: 2021 IEEE
  18th International Symposium on Biomedical Imaging (ISBI). pp. 1764--1767.
  IEEE (2021)

\bibitem{park2019deepsdf}
Park, J.J., Florence, P., Straub, J., Newcombe, R., Lovegrove, S.: Deepsdf:
  Learning continuous signed distance functions for shape representation. In:
  Proceedings of the IEEE/CVF Conference on Computer Vision and Pattern
  Recognition. pp. 165--174 (2019)

\bibitem{Schlegl2019}
Schlegl, T., Seeb{\"o}ck, P., Waldstein, S.M., Langs, G., Schmidt-Erfurth, U.:
  f-anogan: Fast unsupervised anomaly detection with generative adversarial
  networks. Medical image analysis  \textbf{54},  30--44 (2019)

\bibitem{schlegl2017unsupervised}
Schlegl, T., Seeb{\"o}ck, P., Waldstein, S.M., Schmidt-Erfurth, U., Langs, G.:
  Unsupervised anomaly detection with generative adversarial networks to guide
  marker discovery. In: International conference on information processing in
  medical imaging. pp. 146--157. Springer (2017)

\bibitem{Tan2020}
Tan, J., Hou, B., Batten, J., Qiu, H., Kainz, B.: Detecting outliers with
  foreign patch interpolation. arXiv preprint arXiv:2011.04197  (2020)

\bibitem{van2012human}
Van~Essen, D.C., Ugurbil, K., Auerbach, E., Barch, D., Behrens, T.E., Bucholz,
  R., Chang, A., Chen, L., Corbetta, M., Curtiss, S.W., et~al.: The human
  connectome project: a data acquisition perspective. Neuroimage
  \textbf{62}(4),  2222--2231 (2012)

\bibitem{Zimmerer2019}
Zimmerer, D., Isensee, F., Petersen, J., Kohl, S., Maier-Hein, K.: Unsupervised
  anomaly localization using variational auto-encoders. Medical Image Computing
  and Computer Assisted Intervention – MICCAI 2019. Lecture Notes in Computer
  Science  \textbf{11767} (2019)

\bibitem{zimmerer2018context}
Zimmerer, D., Kohl, S.A., Petersen, J., Isensee, F., Maier-Hein, K.H.:
  Context-encoding variational autoencoder for unsupervised anomaly detection.
  arXiv preprint arXiv:1812.05941  (2018)

\bibitem{moodchallenge2020}
Zimmerer, D., Petersen, J., Köhler, G., Jäger, P., Full, P., Roß, T., Adler,
  T., Reinke, A., Maier-Hein, L., Maier-Hein, K.: Medical out-of-distribution
  analysis challenge (Mar 2020). \doi{10.5281/zenodo.3715870},
  \url{https://doi.org/10.5281/zenodo.3715870}

\end{thebibliography}

\end{document}